\documentstyle[aps,prl,preprint]{revtex}
\begin{document}
\title{Structure and energetics of the Si-SiO$_2$ interface}
\author{Yuhai Tu and J. Tersoff}
\address{IBM Research Division, T. J. Watson Research Center\\
P. O. Box 218, Yorktown Heights, NY 10598}

\date{\today}

\maketitle



\vskip 0.5cm

{\bf
Silicon has long been synonymous with semiconductor technology.
This unique role is due largely to the remarkable
properties of the Si-SiO$_2$ interface, especially the
(001)-oriented interface used in most devices\cite{review}.
Although Si is crystalline and the oxide is amorphous,
the interface is essentially perfect, with an extremely
low density of dangling bonds or other electrically active
defects.
With the continual decrease of device size, the nanoscale structure
of the silicon/oxide interface becomes more and more important.
Yet despite its essential role, the atomic structure of
this interface is still unclear.
Using a novel Monte Carlo approach, we identify
low-energy structures for the interface.
The optimal structure found consists
of Si-O-Si ``bridges" ordered in a stripe pattern,
with very low energy.  This structure explains
several puzzling experimental observations.
}

Experiments offer many clues to the interface structure,
but their interpretation remains controversial,
because of the complexities inherent in studying
disordered materials.
Proposed models range 
from a graded interface \cite{HMTYH,Grunthaner} to 
a sharp interface \cite{Luh}
and even to a
crystalline oxide layer
at the interface \cite{Ourmazd}.
Most theoretical studies have involved guessing reasonable
structures \cite{Hybertsen},
sometimes even using hand-built models \cite{handbuilt1}.
More recently, there have been attempts to obtain an unbiased
structure using unconstrained molecular dynamics (MD) \cite{Pasquarello}
and Monte Carlo (MC) studies \cite{Ng}.  However, because of kinetic
limitations these studies have not been able to identify the equilibrium
structure.  Calculations of the interface energy are also not
possible with existing methods.

Here we employ a novel approach in which the Si-SiO$_{2}$
system is modeled as a continuous
network of bonds connecting the atoms,
and the thermodynamic ensemble of possible network topologies
is explored via Monte Carlo (MC) sampling.
The basic method has been described elsewhere \cite{us}
in a simpler context.
But the present application demonstrates that this approach
makes possible an entirely new class of computational studies
of disordered systems.

Our approach samples only defect-free configurations,
in which Si and O have four and two bonds respectively,
and there are no O-O bonds.
Because of this restriction,
the energy may be reasonably approximated by a valence-force model:
\begin{eqnarray}
E_{\{r\}} &=& \frac{1}{2}\sum_{i}k_{b}(b_{i}-b_{0})^{2} \\
&&+\frac{1}{2}\sum_{i,j}k_{\theta }
(\cos \theta _{ij}-\cos \theta _{0})^{2}~. \nonumber
\end{eqnarray}
Here $\{{\bf r}\}$ is the set of atom positions, $E_{\{r\}}$ is the total
energy for a given network topology and given $\{{\bf r}\}$. $i$
represents the $i$th bond, 
$b_{i}$ is ts length,
$\theta _{ij}$ is the angle between bonds $i$ and $j$ 
connected to a common atom.
The material parameters depend implicitly on the type of atom,
where $b_{0}$ is
the preferred bond length, $\theta _{0}$ is the preferred bond angle,
and $k_{\theta }$ and $k_{b}$ are ``spring constants.''
We take
$k_{b,\text{Si-Si}}=9.08 eV/\AA^{2}$,
$k_{\theta,\text{Si-Si-Si}}=3.58eV$,
$b_{0,\text{Si-Si}}=2.35\AA$, $cos(\theta_{0,\text{Si}})=-1/3$,
$k_{b,\text{Si-O}}=27.0eV$, $b_{0,\text{Si-O}}=1.6\AA$,
$k_{\theta,\text{O-Si-O}}=4.32eV$ and $k_{\theta,\text{Si-O-Si}}=0.75eV$,
and $cos(\theta_{0,\text{O}})=-1$.
For Si-Si-O bonds we set the spring constant to be the
geometric mean
$k_{\theta,\text{Si-Si-O}}=(k_{\theta,\text{Si-Si-Si}}%
k_{\theta,\text{O-Si-O}})^{1/2}$.
(There is an additional term in the energy which simply
enforces the restriction of two and four neighbors for
O and Si, respectively \cite{us}.)

In order to focus on the role of network structure, we treat
the energy as a function solely of bond {\it topology},
minimizing $E_{\{r\}}$ with respect to
the geometrical coordinates $\{{\bf r}\}$.
Thus for a given network topology
\begin{equation}
E=\min_{\{{\bf r}\}}E_{\{r\}}~.
\end{equation}
The structure of the system is allowed to evolve toward thermodynamic
equilibrium through Monte-Carlo bond-switching moves \cite{us,www}.
(We adapt the original move to preclude O-O bonds.)
At each step a random trial bond-switch is accepted
with probability $e^{- \Delta E / kT}$
(or 1 for energy-lowering moves), guaranteeing
that the system will evolve toward thermodynamic equilibrium.
This approach gives a fairly accurate description of the structure
of both amorphous Si \cite{us,www} and amorphous SiO$_2$.
Specifically, we have verified that the average bondlength
and bond angles are in agreement with experiment, and
the elastic constants are accurate to better than 20\%.

Our model for the energy is
rather simple compared with the more accurate {\it ab initio} methods
used in some recent studies \cite{Hybertsen,Pasquarello,Ng}.
Since the Si-SiO$_2$ system is dominated by steric constraints,
our approach should nevertheless be reasonably accurate
for the defect-free structures considered here.
More important, it allows the large-scale MC sampling
necessary for the system to move toward thermodynamic
equilibrium, which is not feasible with {\it ab initio} methods.
It also allows us to determine the actual interface energy,
using thermodynamic averaging.  More accurate methods are not
at present able to determine the interface energy,
even for a given interface structure,
because it is impractical to average over the statistical ensemble
of configurations of the amorphous oxide.

We begin with a convenient
though unphysical structure, a perfect interface between
crystalline Si and highly strained $\alpha$-crystobalite.
We use 10 layers of Si, and SiO$_2$ containing an equal number of Si atoms,
periodically repeated in the interface-normal (z) direction.
In the other two dimension, we use cells with
2$\times$2 and 4$\times$4 periodicity, for a total of
160 and 640 atoms respectively.
To accurately describe an oxide at a free Si surface,
the cell size is constrained to match Si(001) in two directions,
while the period normal to the interface is allowed to vary
to maintain zero stress in that direction.

We first perform MC bond-switching within the oxide, allowing it to
amorphize and to relax the large strain by viscous flow.
We then perform unconstrained MC switching of the entire system,
allowing it to equilibrate for up to a total of 300,000 MC steps.
To accelerate the evolution,
the MC ``temperature" used is quite high, 2600$^\circ$C;
but this refers only to the degree of disorder allowed
in the network topology \cite{us}.


We have carried out 10 independent MC simulations for a 2$\times$2 cell.
The resulting interface structures are shown in Fig.~1.
The key structural element is an oxygen bridge between
each pair of Si atoms terminating the Si crystal.
This eliminates half the bonds from the Si side,
correcting the mismatch between the bond densities
in the two very different materials.
This structure allows each atom to maintain its preferred
coordination, with essentially no additional distortion of
the bond angles or bond length beyond that already present due to
the amorphous nature of the oxide.
Bridge bonds have appeared in several previous models of the
Si-SiO$_2$ interface\cite{Ourmazd,Hybertsen,handbuilt1,Ng}.
However, it has apparently not been previously recognized that these
are the key element, giving an ideal low-energy interface.

All 10 simulations gave fully bridge-bonded structures.
However, two distinct arrangements are possible within
our 2$\times$2 periodicity, and both occur in the simulations.
We refer to these as the ``stripe" and ``check" phases,
respectively, and they are compared in Fig.~1.
We have obtained the same structures using a somewhat different
energy function \cite{Laughlin}, showing that it does not
depend on the precise parameter values used.

Similar runs with a 4$\times$4 cell also give bridge-bonded structures,
but the system typically becomes ``stuck"
in a metastable state with incomplete (of order 75\%)
bridge bonding.
The energy is consistently lower in structures with
more complete bridge bonding.
The key role of the bridge bonds is illustrated in Fig.~2,
where the total energy of the system and the number of
bridge bonds at the interface are plotted against MC ``time" for
a typical 4$\times$4 simulation.  There is a clear drop in energy
each time a new bridge bond is formed.
A fully bridge-bonded
structure has the lowest energy and is stable under annealing.
Thus it seems clear that, with sufficiently long annealing,
the 4$\times$4 cell would always reach the ideal stripe structure.
A side view of this structure is shown in Fig.~3a.

In equilibrium, the actual interface structure is that
which minimizes the interface energy (or more precisely,
the free energy).
The interface energy can be calculated
by subtracting the bulk energy of the amorphous oxide and
crystalline Si (obtained in independent
calculations) from the total energy.
In all cases the energy is averaged over
roughly 10,000 MC steps after the system reaches equilibrium.
For the stripe phase,
the calculated interface energy is 6.8$\pm 1.3$ meV/\AA$^2$
(0.10 eV per 1$\times$1 cell), an order of magnitude smaller
than the energy of a free Si surface.
For the check phase, we find a slightly higher energy of
9.5$\pm$1.9 meV/\AA$^2$ (0.14 eV per 1$\times$1 cell).

We can gain further insight into the energetics
by decomposing the total energy of the system
into individual atomic contributions.
This decomposition is not unique, but a natural choice is to
divide the bond-stretching energy in Eq.~1 equally between the
two atoms.
Half of the bond-angle energy is assigned to the vertex atom, and
one quarter to each of the other atoms.

In Fig.~3b, the energy of each atom is plotted versus the
z coordinate, for the low-energy 4$\times$4 stripe structure
after equilibration for 300,000 MC steps.
A striking feature is that the main contribution to the interface energy
comes from local distortions inside the crystalline Si.
The energy within the oxide is rather uniform,
even right up to the interface.

There has been considerable interest in the possibility
of a crystalline interfacial oxide \cite{Ourmazd,Renaud}.
We can form an interface between Si(001) and tridymite (0001)
which resembles the stripe phase above,
but the tridymite is under considerable strain (about 7$\%$ in
one direction and 13$\%$ in the other).
The properties of this interface are summarized in Fig.~4.
The interface energy is actually much higher
than that for amorphous SiO$_2$,
about 29 meV/\AA$^{2}$ (0.43 eV per 1$\times$1 cell).
Thus there appears to be no interfacial driving force
for formation of a crystalline oxide.

Yet several experiments have suggested the presence of
a crystalline oxide layer roughly 5\AA\  thick at the interface,
based on both electron microscopy \cite{Ourmazd}
and x-ray diffraction \cite{Renaud}.
These results have remained an outstanding puzzle,
but they are immediately explained by our structure.

Electron microscopy suggested a 5\AA\  layer of tridymite
at the interface \cite{Ourmazd}.
Comparison of Fig.~3a and Fig.~4a shows that
the structure of the Si-tridymite interface is indistinguishable
from the more realistic crystal-amorphous interface,
in a region several angstroms thick at the interface.
Thus our proposed interface structure is entirely
consistent with the electron microscopy results.
However, it is best viewed as an ordered interface structure,
without reference to any crystalline bulk phase.
In no case did we see evidence for an ordered phase
extending further into the oxide.

X-ray diffraction experiments show an
ordered 2$\times$1 structure at the interface,
with a thickness of under 6 \AA\  and a domain size
comparable to the step spacing \cite{Renaud}.
The stripe phase exactly satisfies these characteristics.
It has an overall 2$\times$1 periodicity. Moreover, every interface step
causes a 90$^\circ$ rotation from 2$\times$1 to 1$\times$2,
so the step spacing sets an upper bound on the domain size.
The presence of random small atomic displacements,
reflected in Figs.~1 and 3, explains the inability of Ref.~\cite{Renaud}
to determine precise atomic positions from the diffraction data.

Finally we note that in several experiments,
photoemission has been used to measure the number
of Si atoms at the interface having intermediate oxidation states
\cite{HMTYH,Grunthaner,holl2}.
Many theoretical studies have attempted to reproduce or explain
these statistics \cite{Hybertsen,Ng},
but the interpretation is surprisingly subtle \cite{holl2}.
Nevertheless, there appears to be some concensus that
the primary connection between Si and SiO$_2$ occurs via
Si$^{\text{+2}}$ \cite{Grunthaner,holl2}, as in our model.

In conclusion, we have identified a simple, defect-free,
ordered structure for the Si-SiO$_2$ interface.
It has low energy, and appears to reconcile the various
puzzling experimental observations.
The computational method focuses
on a more complete exploration of the thermodynamic
ensemble, even when this requires significant approximations
in treating the energetics.
It is our hope that this approach will open the door
to a new class of computation studies of disordered systems.

\newpage
\begin{center}
{\bf Figure Captions}
\end{center}

Fig.~1. Plan view of two Si-SiO$_2$ interfaces.
The last three layers of Si are shown in gold,
with atoms further from the interface shown smaller.
The first layer of O is shown in red.
(a) Stripe phase, having (2$\times$1) symmetry.
(b) Check phase, having c(2$\times$2) symmetry.

Fig.~2. Total energy $E$, and number of interfacial bridge bonds,
versus number of accepted Monte Carlo steps.
The decrease in the energy each time a bridge
bond forms illustrates their crucial role in giving
a low interface energy.

Fig.~3. (a) Side view of 4$\times$4 stripe phase ([110] projection). The
Si and O atoms are represented by gold and red spheres respectively.
Each arrow points to a row of oxygen atoms that form the bridges at
the interface. Notice the
substantial voids above each bridge bond.
(b) Energy of each atom versus its z coordinate.
Red circles represent oxygen atoms and gold circles
represent silicon atoms.
The green line is the local energy per atom, averaged
over 20 configurations
(and over a z range of $\sim 1\AA$ for smoothness).

Fig.~4. Same as Fig.~3, for interface between
Si and tridymite.

\end{document}